# Bauschinger Effect in Thin Metal Films: Discrete Dislocation Dynamics Study

Kamyar M. Davoudi [a], Lucia Nicola [b], Joost J. Vlassak [a*]

[a] School of Engineering and Applied Sciences, Harvard University, Cambridge, MA 02138, USA

[b] Department of Materials Science and Engineering, Delft University of Technology, Mekelweg 2, 2628 CD Delft, The Netherlands

**Abstract**

The effects of dislocation climb on plastic deformation during loading and unloading are studied using a two-dimensional discrete dislocation dynamics model. Simulations are performed for polycrystalline thin films passivated on both surfaces. Dislocation climb lowers the overall level of the stress inside thin films and reduces the work hardening rate. Climb decreases the density of dislocations in pile-ups and reduces back stresses. These factors result in a smaller Bauschinger effect on unloading compared to simulations without climb. As dislocations continue to climb at the onset of unloading and the dislocation density continues to increase, the initial unloading slope increases with decreasing unloading rate. Because climb disperses dislocations, fewer dislocations are annihilated during unloading, leading to a higher dislocation density at the end of the unloading step.

---

[*] Author to whom correspondence should be addressed. Electronic mail: vlassak@seas.harvard.edu

---

## I. Introduction

The use of metal thin films in micro-electronic and micro-electro-mechanical systems (MEMS) has motivated research on their mechanical properties. Thin films on substrates often experience temperature cycles that lead to plastic deformation because of differential thermal expansion between film and substrate. As plastic deformation directly impacts the level of stress in a film and hence its reliability, there has been a strong drive to study plasticity in thin films. The structure of thin films also makes them excellent vehicles to probe fundamental problems in materials science: The grain size of a thin film is often much smaller than that of the same material at the macro-scale and grains are often columnar. The resulting proximity of free surfaces and the high density of interfaces in thin films have a profound impact on their mechanical behavior that is not yet fully understood.[1]

In general, thin films can support much higher stresses than their bulk counterparts, and their mechanical response is size dependent. Because the constitutive equations in classical continuum theories do not have internal length scales, these theories cannot predict size-dependent responses. There have been a considerable number of attempts to develop continuum theories that incorporate one or more length scales into the constitutive equations, including nonlocal [2] and strain-gradient theories.[3-9] Despite these attempts, there is no continuum theory that can predict the behavior of materials in all experiments.

It is well known that in crystalline solids plasticity at small scales takes place by the same fundamental mechanisms observed in bulk materials: plastic flow proceeds mainly by the collective motion of dislocations. This observation affords the use of discrete dislocation dynamics (DDD) to study plasticity in thin films. In the DDD approach, dislocations are modeled as line singularities in an isotropic, elastic solid. The behavior of the dislocations is governed by a set of simple constitutive equations that describe how they move, nucleate, and interact with obstacles. Although three-dimensional DDD models capture the physics of problems more accurately than two-dimensional models, they are computationally demanding and are not easily applied to thin films. Therefore, most three-dimensional models are limited to single crystals, very small strains, small volumes of material, and low dislocation densities. For example, `ParaDis`, a powerful three-dimensional DDD code that was originally developed at the Lawrence Livermore National Laboratory[10], can only model single-crystal materials. Two-dimensional discrete dislocation dynamics models, on the other hand, can model polycrystalline materials, realistic dislocation densities, and relatively large strains with much less computational effort.

If a metal is deformed plastically in one direction, plastic deformation often starts at a much lower stress level upon reversal of the load, a phenomenon known as the Bauschinger effect. Departure from the linear unloading curve during reverse deformation sometimes begins before the stress changes sign. The Bauschinger effect is a natural consequence of the inhomogeneous nature of plastic flow; understanding the fundamental causes of the effect is an essential step towards developing better strain hardening theories and constitutive models for cyclic deformation[11]. The Bauschinger effect is generally ascribed to either short-range effects, such as the directionality of mobile dislocations in their resistance to motion or the annihilation of dislocations during reverse loading, or to long-range effects such as back stresses caused by dislocation pile-ups at grain boundaries or obstacles. Both effects assist plastic deformation in the reverse direction and can give rise to a Bauschinger effect.[12]

Xiang and Vlassak[13,14] reported the first direct observations of the Bauschinger effect in thin films. They found that the effect could be quite significant in thin films, especially if the films were passivated. Their findings were explained as the result of large back stresses caused by dislocation pile-ups at the passivation layers. The Bauschinger effect in thin films was modeled using two-dimensional[15,16] and 2.5-dimensional[17] DDD simulations. These simple two-dimensional models captured the size dependence of the yield stress of thin films [15], but they overestimated the stresses caused by work hardening because they lacked any softening or recovery mechanisms. Recently, dislocation climb was introduced to two-dimensional DDD by Davoudi et al. [18] for polycrystalline thin films and by Deshpande et al.[19,20] for single crystals. In this paper, we use discrete dislocation dynamics to investigate the Bauschinger effect in polycrystalline thin films. The analyses have been carried out using a two-dimensional DDD model that includes dislocation climb to better describe strain-hardening behavior. While the analyses focus on the effects of dislocation climb, climb may be taken as representative of a range of softening mechanisms that occur in a material.

## II.  Discrete Dislocation Dynamics Framework

In discrete dislocation dynamics, a material is generally modeled as an elastic solid containing dislocations. As a load is applied to the material, the dislocations are allowed to move and evolve incrementally. At any instant in time, it is assumed that the material is in equilibrium and that the displacement and stress fields are known. An increment of strain is prescribed and the positions of the dislocations, the displacement field, and the stress field are updated using the following procedure: (1) The Peach-Koehler force is calculated along the length of each dislocation; (2) the dislocation structure is allowed to evolve in response to the Peach-Koehler force by a number of mechanisms including dislocation nucleation, motion, and annihilation; (3) the stress state in the solid is calculated for the updated dislocation arrangement. Steps 1 and 3 follow from elasticity; step 2 requires the formulation of constitutive rules for dislocation behavior. In this paper, we follow the rules suggested by Kubin et al.[21] for dislocation glide, dislocation annihilation and dislocation nucleation.

Determining the stress state at each time step requires the solution of an elastic boundary value problem. In the two-dimensional DDD framework developed by Van der Giessen and Needleman,[22] the displacement, strain and stress fields are written as the superposition of two fields,

$$u = \tilde{u} + \hat{u}, \quad \varepsilon = \tilde{\varepsilon} + \hat{\varepsilon}, \quad \sigma = \tilde{\sigma} + \hat{\sigma}. \tag{1}$$

The (~) fields are obtained by summing the fields associated with the individual dislocations in the material under the assumption of an infinite medium,

$$\tilde{u} = \sum_{I=1}^{N} u^{(I)}, \quad \tilde{\varepsilon} = \sum_{I=1}^{N} \varepsilon^{(I)}, \quad \tilde{\sigma} = \sum_{I=1}^{N} \sigma^{(I)}, \tag{2}$$

where $u^{(I)}$, $\varepsilon^{(I)}$, and $\sigma^{(I)}$ are the fields due to dislocation $I$, analytical expressions for which can be found in standard texts (see, e.g., Ref. [23]). The (^) fields represent the image fields that enforce the correct boundary conditions. They are smooth and are readily calculated using the finite element method or a boundary element analysis. The Peach-Koehler force on a dislocation $I$ is given by

$$F^{(I)} = \left[\left(\hat{\sigma} + \sum_{J \neq I} \sigma^{(J)}\right) \cdot b\right] \times \xi, \tag{3}$$

where $\xi$ is the local tangent to the dislocation line and $b$ is the Burgers vector. The glide component of this force is $F_g^{(I)} = F^{(I)} \cdot (\xi \times n)$ and the climb component $F_c^{(I)} = F^{(I)} \cdot n$, where $n = b \times \xi / \|b \times \xi\|$ is the unit vector perpendicular to the glide plane of the dislocation.

Simulations typically start with the material in a dislocation-free state. Dislocation sources are randomly distributed on the slip planes with each source characterized by nucleation strength, $t_{\text{nuc}}$. When the glide component of the Peach-Koehler force on a dislocation source exceeds $b\tau_{nuc}$ during a time $t_{nuc}$, two dislocations of opposite sign are nucleated on the glide plane. The distance between the newly formed dislocations,

$$L_{nuc} = \frac{\mu}{2\pi(1-\nu)} \frac{b}{\tau_{nuc}}, \qquad (4)$$

is taken such that the attraction between the two dislocations is balanced by $t_{nuc}$, where $\mu$ is the shear modulus and $n$ is Poisson's ratio of the material. When two dislocations of opposite sign come closer to each other than a critical distance $L_{ann}$, they annihilate each other and are removed from the model. According to experimental [24] and computational evidence [25], the glide velocity in an fcc material without internal obstacles is a linear function of the glide force. This is also the relationship used in this DDD model, i.e., $V_g^{(I)} = F_g^{(I)}/B$, where $B$ is called the drag coefficient, a quantity that increases linearly with temperature.[24]

Dislocation climb is implemented in the DDD simulations using the following model [18]. Consider a dislocation as a perfect source or sink of vacancies at the center of a cylinder of radius $R$, and take the equilibrium concentration of vacancies in the cylinder to be $c_0$. When a force $F_c$, is suddenly applied to the dislocation in the direction perpendicular to the dislocation glide plane, the dislocation starts to climb, absorbing or emitting vacancies until a concentration of $c = c_0 \exp(-F_c b^2/k_B T)$ is reached near the dislocation core. In this expression $k_B$ is the Boltzmann constant, $T$ refers to the absolute temperature, and $b$ is the magnitude of the Burgers vector of the dislocation. At that vacancy concentration, the chemical force due to the departure from the equilibrium concentration balances the mechanical force $F_c$. As a result of the ensuing gradient in chemical potential, there is a diffusive flux of vacancies, which determines the rate of climb. Assuming steady-state diffusion inside the cylinder and further assuming that the concentration at a distance $R$ remains $c_0$, the climb velocity is given by [26–28]

$$V_c = \frac{2\pi D_0}{b \ln(R/b)} \exp\left(-\frac{\Delta E_{sd}}{k_B T}\right) \left[\exp\left(\frac{F_c b^2}{k_B T}\right) - 1\right],  \quad (5)$$

where $\Delta E_{sd}$ is the vacancy self-diffusion energy, and $D_0$ the pre-exponential diffusion constant. The climb force is taken positive when it favors vacancy emission.

At each time step, the glide and climb velocities of the dislocations in the simulation are calculated and the positions of the dislocations are updated accordingly. Because the climb velocity is typically much smaller than the glide velocity, different time steps are used for climb and glide. In this paper, the time step for climb is taken 100 times larger than the time step for glide.

When one of the dislocations in a dislocation dipole climbs out of its original glide plane, simple superposition of the individual displacement fields of these two dislocations does not provide the correct discontinuity in the displacement field of the non-planar dislocation dipole. To overcome this shortcoming and to find the correct displacement field due to a dislocation dipole where one of the dislocations climbs from $(x_0,y_0)$ to $(x_0,y_1)$, the following terms need to be added to the $x$-component of the displacement field published in most texts on dislocations, for example Eq. (2.15) of Ref. 29.

$$\frac{b}{2\pi}\left[\tan^{-1}\left(\frac{y-y_1}{x-x_0}\right) - \tan^{-1}\left(\frac{y-y_0}{x-x_0}\right) + \tan^{-1}\left(\frac{x-x_0}{y-y_1}\right) - \tan^{-1}\left(\frac{x-x_0}{y-y_0}\right)\right]. \tag{6}$$

These extra terms account for the displacement caused by the emission or absorption of vacancies during climb[18].

## III. Thin Film Model and Selection of Parameters

Simulations were carried out on freestanding polycrystalline films passivated on both surfaces. The films were subjected to plane-strain tension as illustrated schematically in Fig. 1. In line with Nicola et al.[15], the film was modeled as a two-dimensional array of rectangular grains of thickness $h$. In doing so, a periodic unit-cell of width $w$ consisting of six randomly oriented grains of uniform size $d$ was considered. Plane-strain conditions were assumed normal to the $xy$-plane. Grain boundaries and passivation layers were assumed impenetrable to dislocations. Each grain had three sets of slip planes that differed by an angle of 60°.[30] As mentioned earlier, the grains were initially dislocation free, but Frank-Read sources were distributed randomly on the slip planes in the grains. No obstacles were present to impede dislocation motion. Tension was imposed by prescribing a constant displacement rate difference between the left and right edges of the unit-cell. The top and bottom surfaces of the unit-cell were taken to be traction-free. The average stress in the film, $s$, is calculated as

$$\sigma = \frac{1}{h}\int_h \sigma_{xx}(w,y)\mathrm{d}y, \tag{7}$$

where the integral over the film thickness excludes the passivation layers.

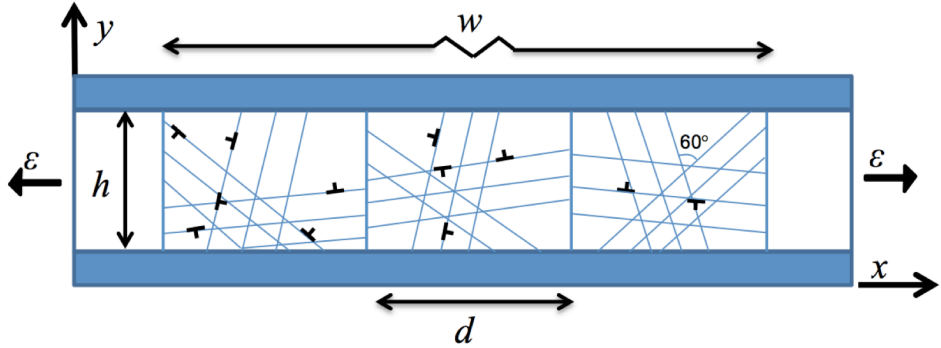

FIG. 1: Schematic representation of the thin film model

The physical properties that were used for the film material are representative of aluminum and are given in Table I. The passivation layers were assumed to remain elastic and had the same elastic properties as the film material. The thickness of the film and the passivation layers on both film surfaces were 750 nm and 20 nm, respectively; the grain size was taken as 1.0 µm. All simulations were run at a temperature of 900 K. The drag coefficient was taken as $3.2 \times 10^{-5}$ Pa s.[18] The annihilation distance $L_{\text{ann}}$ and the nucleation time $t_{\text{nuc}}$ were chosen as $6b$ and 10 ns, respectively. The density of dislocation sources was 15 µm$^{-2}$ in all simulations. The source strength $t_{\text{nuc}}$ was randomly chosen from a Gaussian distribution of strengths with an average value of 100 MPa and a standard deviation of 20 MPa. The values of these parameters were estimated by fitting simulated curves to experimental stress-strain curves for thin films deformed under tension at room temperature[15]. Because climb allows dislocations to leave their glide planes, dislocations can occur on all possible glide planes in the material, not just those with dislocation sources. The spacing between glide planes was taken equal to $b$ in all simulations. To limit computation time, a high strain rate $|\dot{\varepsilon}| = 4000$ s$^{-1}$ was used for all simulations except otherwise indicated; the time step was taken to be 0.5 ns. To reduce the statistical effects of the initial conditions, at least four realizations of the model that differed from each other with respect to the locations of the dislocation sources were run for each set of parameters.

| Parameter | Value | Parameter | Value |
|---|---|---|---|
| $\mu$ | 26 GPa | $\Delta E_{sd}$ | 1.28 eV |
| $\nu$ | 0.35 | $D_0$ | 0.1185 cm$^2$/s |
| $b$ | 2.86 Å | | |

**Table I:** Materials properties taken in the simulations

## IV. Results and Discussion

Figure 2(a) shows two stress-strain curves for a 750 nm film passivated on both surfaces, one curve for the case where dislocations are allowed to glide and one curve where they can both glide and climb. The dashed lines represent linear elastic unloading and have slopes given by the plane-strain modulus of the film, $E/(1-\nu^2)$. Figure 2(b) is the same as Fig. 2(a), but here the stress is plotted against the plastic strain. The vertical dashed lines represent elastic unloading. It is evident from the figures that the strain-hardening rate is much reduced if dislocations are allowed to climb out of their glide planes. This behavior is of course consistent with the notion that climb is a softening mechanism that results in a more realistic simulation of work hardening.[15,18] Two more features are noteworthy: (1) the stress-strain curves show a significant Bauschinger effect that increases with increasing strain and (2) forward plastic flow continues during initial unloading when climb is allowed.

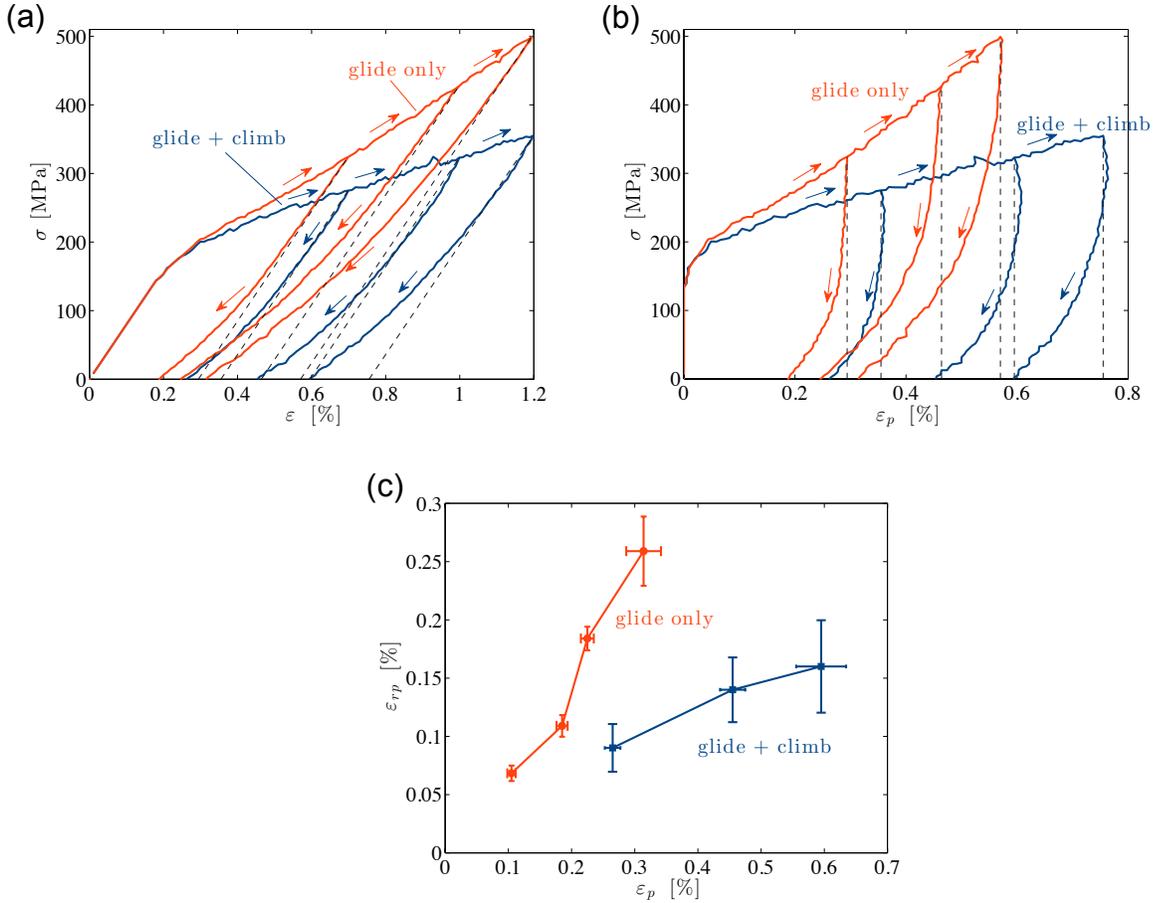

FIG. 2: The average stress as a function of (a) the applied strain (b) the plastic strain for the case of glide only and glide with climb. The dashed curves show the fully elastic unloading. (c) The Bauschinger strain versus the plastic strain in the film either dislocation climb is enabled or disabled. Error bars represent the standard error.

We define the Bauschinger strain, $\varepsilon_{rp}$, as the difference between the actual unloading strain and the elastic unloading strain. Figure 2(c) shows $\varepsilon_{rp}$ as a function of the plastic strain in the film. Evidently dislocation climb reduces the Bauschinger strain significantly. As the deformation proceeds and the stress in the film increases, more dislocation pile-ups are formed, back stresses increase, and the Bauschinger effect becomes more pronounced. Climb allows some dislocations in areas with high stresses such as the tips of dislocation pile-ups to leave their glide planes. This process reduces the back stress on the other dislocations in the pile up and on any dislocation source in that glide plane, as illustrated in Fig. 3. On unloading, the lower back stress reduces the magnitude of the Bauschinger effect compared to the case where dislocations can only glide. Since the Bauschinger effect is induced by back stresses and back stresses are proportional to the density of dislocations in pile-ups, the results in Figure 2(c) suggest that the total density of dislocations in pile-ups should be smaller when climb is allowed. Figure 4 illustrates that this is indeed the case: there is a significant drop in the density of dislocations that are part of a pile-up when dislocations are allowed to climb. The observation that climb reduces the Bauschinger strain, should be contrasted with a recent finding by Deshpande et al. that the Bauschinger effect in single crystals with permeable passivation layers is more pronounced for climb-assisted glide than for glide only, even though climb reduces back stresses.[19] They attribute this apparent contradiction to the permeability of the passivation: Climb-assisted glide results in shorter pile-ups with dislocations spreading themselves more evenly over the film. As a result, the stress acting on the leading dislocations in a pile up is smaller and fewer dislocations can penetrate the passivation layers. Fewer dislocations exit the film and the stored dislocation density is greater than in the absence of climb. This increased dislocation storage enhances the Bauschinger effect. This explanation does not, however, hold for impenetrable passivation layers, for which they also report enhanced dislocation densities in the case of climb-assisted glide. The passivation layers in this study are impenetrable, and an enhanced dislocation density is not observed here.

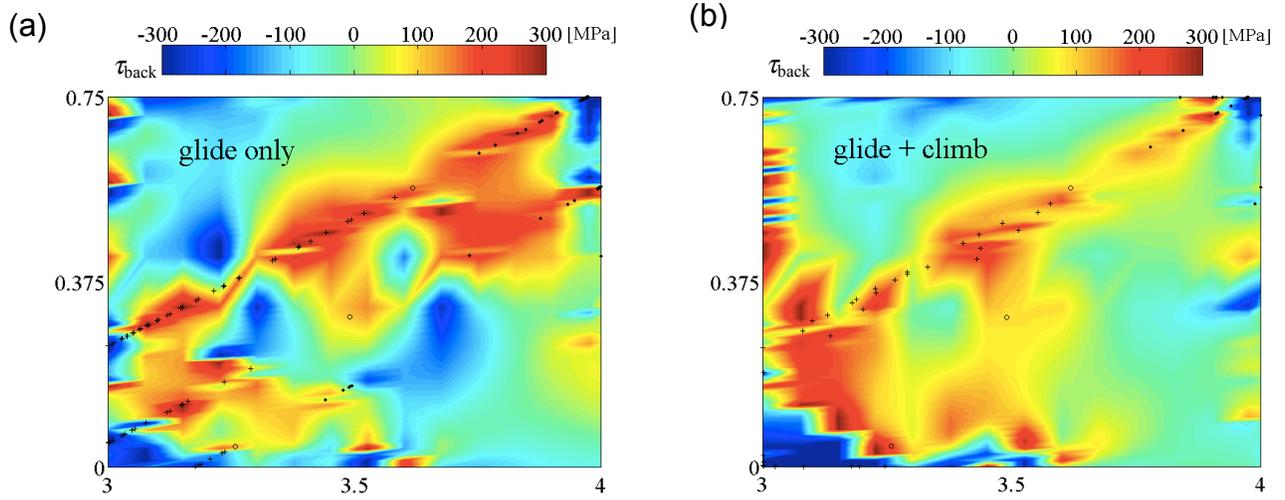

**FIG 3:** Distribution of the back stress on a slip system in a single grain at $\varepsilon = 1.2\%$ for the case of (a) glide only, and (b) glide and climb. The back stress is defined as the difference between the local shear stress and the applied normal stress resolved on a given glide plane. Only dislocations and sources on one set of glide planes are displayed. Positive and negative dislocations are depicted by the "+" and "." symbols, respectively. Open circles denote dislocation sources. The unit of length in the figure is 1 μm.

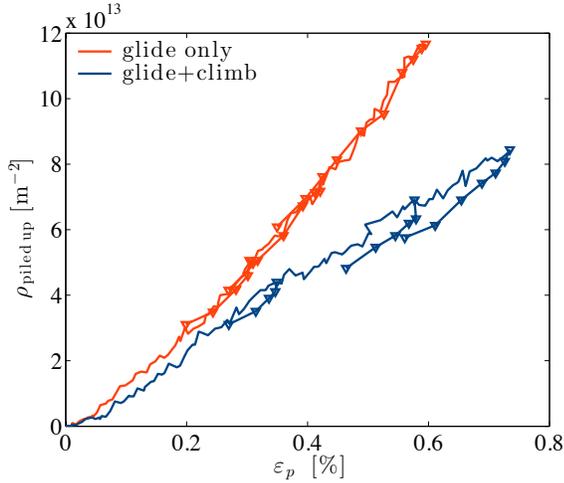

**FIG. 4: Total density of dislocations in pile-ups versus plastic strain for the case of glide only and glide with climb. The markers indicate the unloading curves.**

From the stress-strain curves in Figure 2(b) it is evident that forward plastic flow continues for a while, during unloading when dislocations can climb. This feature becomes more noticeable at slower unloading rates. To illustrate the effect of unloading rate, we have plotted the film stress versus the plastic strain for three different unloading rates in Fig. 5. Because the change in the stress-strain curve is negligible as the loading rate is reduced from 4,000 s$^{-1}$ to 400 s$^{-1}$, only one loading curve is shown in Fig. 5.

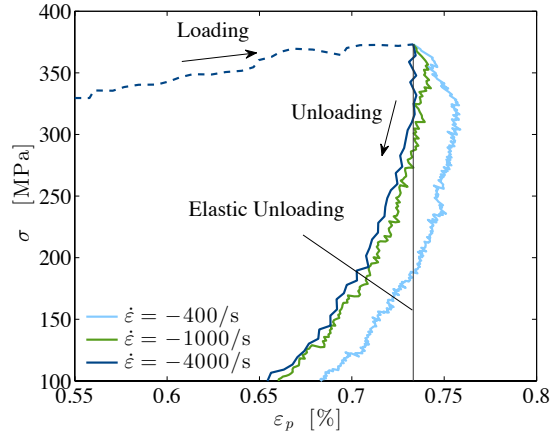

FIG. 5: Stress vs. plastic strain for different unloading rates. The dashed curve and the vertical line show the loading segment where $\dot{\varepsilon} = 4,000$ s$^{-1}$ and the elastic unloading, respectively.

In the limit of a zero unloading rate such as in a stress-relaxation-experiment, the stress decreases in proportion to the creep-induced plastic strain rate and the slope of the unloading curve in Fig. 5 approaches the plane-strain modulus $E/(1-\nu^2)$. Conversely, if unloading happens infinitely fast, dislocations do not have time to move, the process is entirely elastic, and the unloading curve in Fig. 5 has an infinite slope. Therefore, the beginning of any unloading curve should lie between these two limiting cases. When the motion of dislocations is limited to glide, slower loading or unloading rates have a negligible effect on the stress-plastic strain curves. To get more insight in this behavior, the cumulative distance, $L_c$, swept by climbing dislocations is shown as a function of plastic strain in Fig. 6(a). As expected, the figure shows a gradual increase in the cumulative distance during loading. When unloading starts, however, dislocations continue to climb and $L_c$ continues to increase, albeit at a much-reduced rate. Dislocation climb does not lead to an immediate build up of back stresses that shut down the climb process and dislocations climb at a rate commensurate with the local stress, even on unloading. The smaller the unloading rate, the longer the unloading process and the greater the distance swept by climbing dislocations. The connection between the climb distance and the forward plastic flow on unloading is then made via Orowan's equation, which links the plastic strain to the dislocation motion: since $L_c$ continues to grow during initial unloading, so does the plastic strain. In addition to dislocation climb, an increase in dislocation density also contributes to the forward plastic flow on unloading. This point is illustrated in Fig. 6(b), which shows a small increase in total dislocation density during initial unloading – the slower the unloading, the greater the increase. The increase in dislocation density is again a direct consequence of climb: As dislocations climb out of their glide planes at the onset of unloading, the back stresses on the dislocation sources decrease allowing them to emit more dislocations.

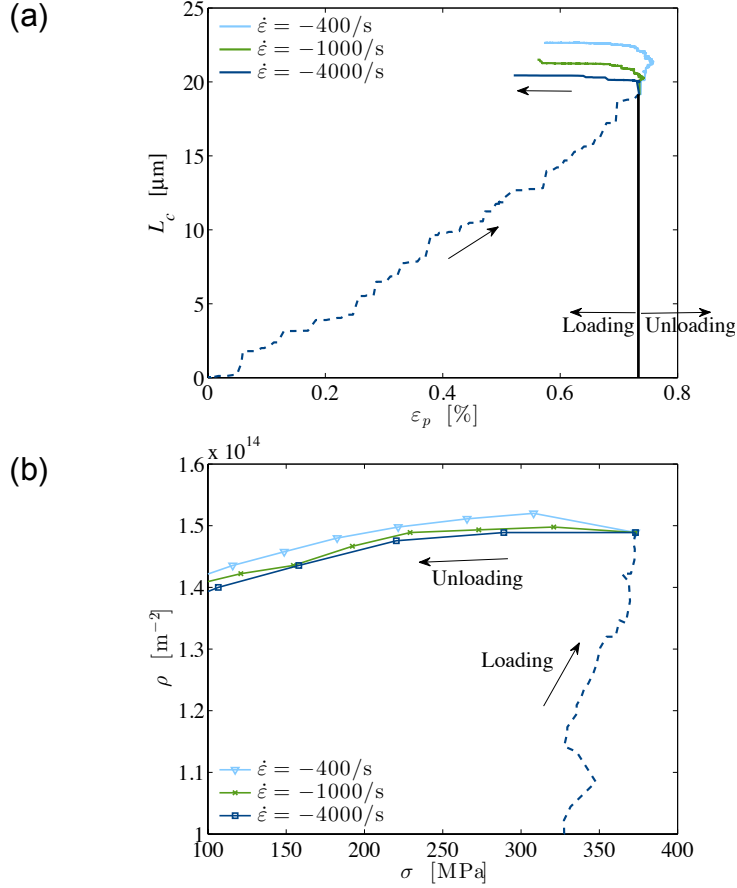

**FIG. 6:** (a) Cumulative distance swept by dislocation climb as a function of plastic strain; (b) total dislocation density versus the plastic strain. The dashed curves show the loading segment, where $\dot{\varepsilon} = 4,000$ s$^{-1}$.

Considering the changes in $L_c$ and $\rho$ during unloading (Fig. 6, 7), the unloading segment of the stress-strain curves in Fig. 2(a) is now readily explained. In general, we have that $\dot{\sigma} = E/(1-v^2)(\dot{\varepsilon} - \dot{\varepsilon}_p)$, where dots indicate incremental changes. At the onset of unloading, both $L_c$ and $r$ increase, $\dot{\varepsilon}_p > 0$ and $d\sigma/d\varepsilon > E/(1-v^2)$. As unloading proceeds, $L_c$ approaches a constant value, while $r$ decreases very slowly; the plastic strain rate is very small, $\dot{\varepsilon}_p \approx 0$ and $d\sigma/d\varepsilon \approx E/(1-v^2)$. Toward the end of the unloading process, dislocations reverse their direction of glide because of back stresses and start to annihilate each other. The dislocation density decreases more rapidly, $\dot{\varepsilon}_p$ becomes negative, and the unloading slope decreases steadily until eventually it becomes smaller than the elastic slope, leading to the Bauschinger effect.

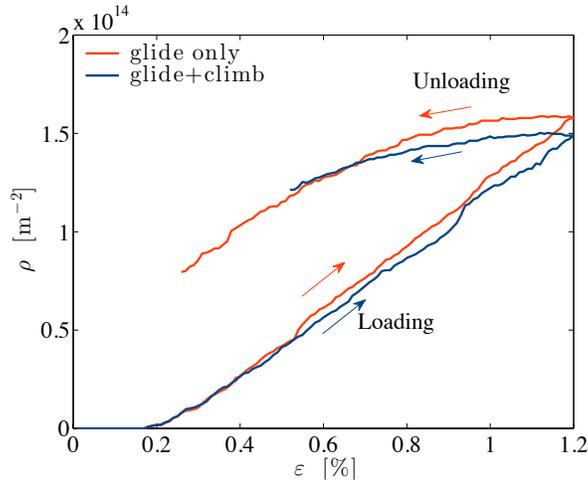

**FIG 7:** Dislocation density versus the applied strain when the loading and unloading rates are 4,000 s$^{-1}$.

In Fig. 7, which plots the dislocation density as a function of applied strain, two observations are worth noting: (1) Although the film is initially dislocation free, many dislocations still exist in the film for both cases when the average stress in the film is reduced to zero. These dislocations remain in the film because stresses induced by other dislocations prevent them from going back even in the presence of back stresses. Furthermore the model lacks line tension, which normally provides a driving force for dislocation loops to collapse, and would be expected to overestimate the number of dislocations that remain in the two-dimensional model. If line tension were incorporated in the model, a more pronounced

Bauschinger effect would be observed. (2) Because dislocations become more dispersed when climb is allowed, the dislocation density decreases less during unloading compared to the glide only case. One would expect this trend to be even more pronounced when switching to a three-dimensional discrete dislocation model, as there are many more mechanisms in a three-dimensional model by which dislocations can be retained in the material.

## V. Conclusions

We have evaluated the effect of dislocation climb on the unloading behavior of thin films using two-dimensional discrete dislocation simulations. Unloading curves obtained in discrete dislocation simulations often have a strong Bauschinger effect. Because dislocation climb results in a more dispersed distribution of dislocations in the film, the total density of dislocations in pile-ups and the magnitude of the back stresses are reduced. As a result, the Bauschinger effect will be less pronounced if the dislocation climb is allowed. At the onset of unloading, dislocations keep climbing, and the dislocation density initially increases, resulting in forward plastic flow during initial unloading, an effect especially pronounced at slow unloading rates. As the unloading process continues, dislocations start to move in the reverse direction and the slope of the stress-strain curve continuously decreases.

## Acknowledgement

The authors gratefully acknowledge support from the National Science Foundation (Grant No. DMR-0906892).